\documentclass[fleqn,10pt,twocolumn]{wlscirep}
\usepackage[utf8]{inputenc}
\usepackage[T1]{fontenc}

\usepackage{lineno}

\usepackage{amsmath,amssymb,amsfonts}
\usepackage{algorithmic}
\usepackage{textcomp}
\usepackage{color,xcolor}
\usepackage{xspace}
\usepackage{longtable}
\usepackage{multirow}
\usepackage{enumitem}
\usepackage{booktabs}
\usepackage{wrapfig}
\usepackage{lipsum}
\usepackage{soulutf8}
\usepackage{scicite}
\usepackage{colortbl}
\usepackage{subcaption}

\makeatletter
\renewcommand{\fnum@figure}{Fig. \thefigure}
\makeatother
\usepackage{enumitem}

\usepackage{amsmath}
\usepackage{amsfonts}
\usepackage{url}

\usepackage{multicol}

\usepackage{csquotes}

\usepackage{afterpage}
\usepackage[ruled,vlined]{algorithm2e}
\usepackage{setspace}

\usepackage{tabularx}  
\usepackage{booktabs}  
\usepackage{multirow}  
\usepackage{caption}   
\usepackage{threeparttable}

\usepackage[style=nature]{biblatex}
\renewbibmacro*{finentry}{%
  \finentry
  \iffieldundef{annotation}
    {}
    {\par\vspace{0.5em}\noindent\textbf{\printfield{annotation}}\par}%
}

\usepackage{pifont}

\usepackage{hyperref}
\hypersetup{
    colorlinks=true,
    linkcolor=magenta,
    filecolor=magenta,      
    urlcolor=blue,
}

\usepackage{xcolor}
\usepackage{soul}
\usepackage{wrapfig}
\usepackage[braket, qm]{qcircuit}
\usepackage{graphicx}

\usepackage{qcircuit}
\usepackage{graphicx}

\usepackage{tikz}
\usetikzlibrary{shapes, arrows.meta, positioning}
\usepackage[most]{tcolorbox}

\usepackage{enumitem} 
\setlist[enumerate]{wide = 0pt, leftmargin=*}
\usepackage{cuted}
\usepackage{stfloats}
\usepackage{mdframed}
\usepackage{comment}

\usepackage[normalem]{ulem}

\newcounter{boxcounter}
\title{Quantum-machine-assisted Drug Discovery}
\author[1]{Yidong Zhou$^\dag$}
\author[2]{Jintai Chen$^\dag$}
\author[3]{Jinglei Cheng}
\author[2]{Xu Cao}
\author[4]{Yuanyuan Zhang}
\author[5]{Gopal Karemore}
\author[6]{Marinka Zitnik}
\author[7]{Frederic T. Chong}
\author[3]{Junyu Liu$^*$}
\author[1]{Tianfan Fu$^*$}
\author[1]{Zhiding Liang$^*$}

\affil[1]{Rensselaer Polytechnic Institute, Troy, NY, USA.}
\affil[2]{University of Illinois at Urbana-Champaign, IL, USA.}
\affil[3]{University of Pittsburgh, Pittsburgh, PA, USA.}
\affil[4]{Purdue University, USA.}
\affil[5]{Novo Nordisk A/S, Måløv, Denmark.}
\affil[6]{Harvard University, Cambridge, MA, USA.}
\affil[7]{University of Chicago, Chicago, IL, USA.}
\affil[$^\dag$]{These authors contribute equally.}
\affil[$^*$]{These authors are co-corresponding authors. Corresponding to junyuliucaltech@gmail.com, futianfan@gmail.com, zlianghahaha@gmail.com}

\begin{abstract}
Drug discovery is lengthy and expensive, with traditional computer-aided design facing limits.  This paper examines integrating quantum computing across the drug development cycle to accelerate and enhance workflows and rigorous decision-making. It highlights quantum approaches for molecular simulation, drug-target interaction prediction, and optimizing clinical trials. Leveraging quantum capabilities could accelerate timelines and costs for bringing therapies to market, improving efficiency and ultimately benefiting public health.
\end{abstract}

\begin{document}

\flushbottom
\maketitle

Drug discovery and development is a highly complex and costly endeavor, typically requiring over a decade and billions of dollars to bring a single therapeutic to market~\cite{huang2022artificial,bohacek1996art}. Traditional computer-aided drug design has made significant strides in accelerating this process, but faces fundamental limitations~\cite{sadybekov2023computational,trajanoska2023target, reker2019computational, liu2021accelerated, chudasama2016recent, zhang2024geometricdeeplearningstructurebased, powers2023geometric, bhardwaj2016accurate, liu2023deep}. The chemical space of potential drug compounds—estimated at $10^{60}$ molecules—vastly exceeds what classical algorithms can efficiently explore, while conventional simulation methods struggle to accurately model the quantum-mechanical interactions that govern molecular behavior~\cite{santagati2024drug}. These challenges, coupled with the multidimensional complexity of clinical trial data and stringent privacy requirements~\cite{li2024hybrid}, create computational bottlenecks throughout the pharmaceutical pipeline that impact both time-to-market and development success rates. The integration of quantum computing across the entire pharmaceutical value chain—from molecular design to clinical trial optimization—represents a particularly promising frontier at this technological intersection.

In this manuscript, we review the quantum computing approaches for drug discovery and development and explore how quantum technologies might transform pharmaceutical innovation. We examine quantum approaches to molecular simulation and lead identification and optimization where native quantum state representation offers exponential advantages over classical methods. We analyze quantum-enhanced clinical trial optimization through advanced resource allocation algorithms and secure data integration frameworks. By systematically comparing computational requirements across application areas, we highlight near-term opportunities for quantum advantage while mapping the trajectory toward fault-tolerant implementations. The convergence of these transformative technologies promises to significantly reduce the time and cost associated with bringing new therapeutics to patients, potentially revolutionizing how we address global health challenges.

\begin{figure}[t!]
    \centering
    \includegraphics[width=\linewidth]{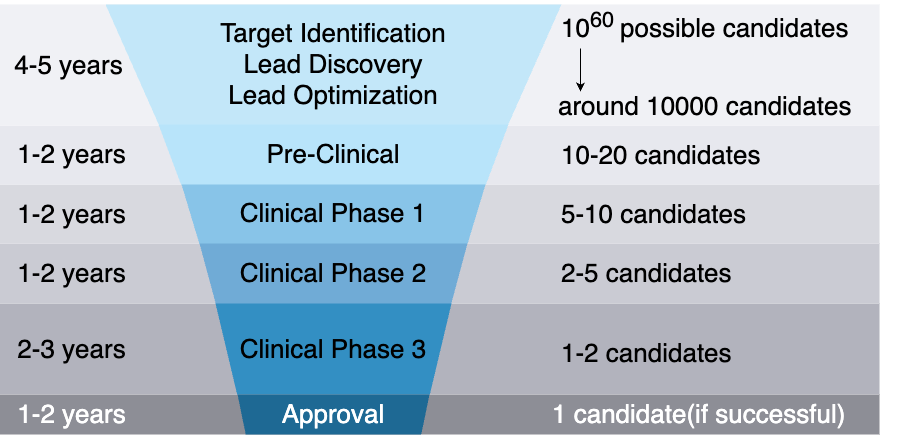}
    \caption{\textbf{Pipeline of drug discovery and development.} Drug discovery prioritizes a small set of novel structures with desirable properties, followed by pre-clinical studies in vivo and phased clinical development to evaluate safety and efficacy. The stage durations and attrition rates illustrated here reflect representative ranges from regulatory and research bodies such as the U.S. Food and Drug Administration (FDA) and the National Institutes of Health (NIH)~\cite{fda_devpath, fda_pdufa_goals, nih_ncats_timeline, ctgov_stats}. The magnitude of small-molecule chemical space and the early-stage funnel from very large libraries to tractable candidate sets are supported by peer-reviewed surveys and modern ultra-large screening exemples~\cite{polishchuk2013estimation, tingle2023zinc, lyu2019ultra, gorgulla2020open, zhou2024artificial}.}
    \label{fig:drugpipline}
\end{figure}

\begin{figure*}[t]
    \centering
    \includegraphics[width=1\textwidth]{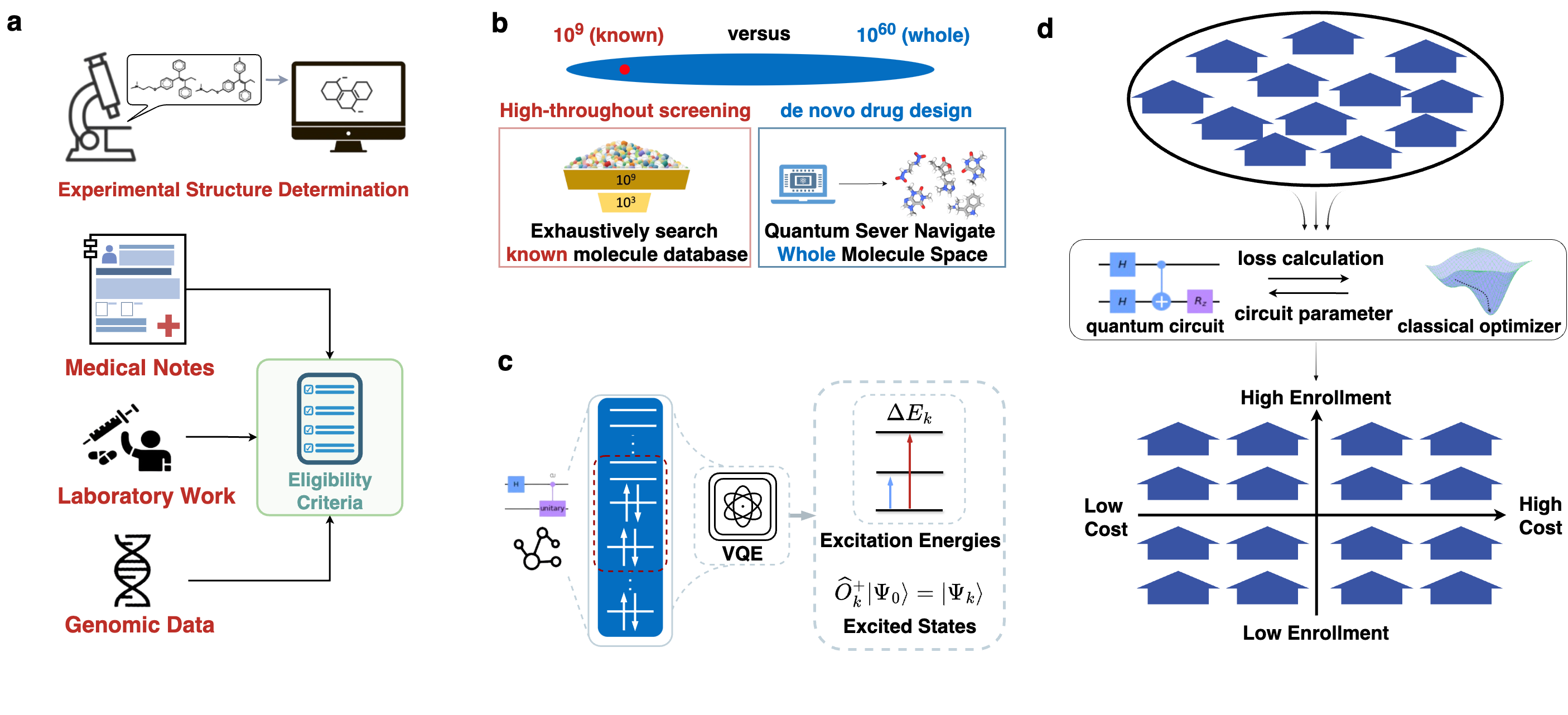}
    \caption{\textbf{Overview of quantum computing integration in drug development and clinical trial optimization.} (a): Classical pre-processing pipeline integrating diverse data sources for eligibility criteria determination, including experimental structure analysis via microscopy, patient medical records, laboratory test results, and genomic data. (b): Comparative analysis of classical versus quantum approaches in drug discovery, contrasting traditional high-throughput screening methods that search through approximately $10^9$ known molecules with quantum-enabled \textit{de novo} drug design capable of exploring a vastly larger chemical space of $10^{60}$ molecules. The quantum approach uses quantum algorithms to navigate this expanded molecular space more efficiently than classical methods. (c): Detailed quantum algorithm implementation showing the progression from molecular structure to quantum circuit design, incorporating VQE for calculating molecular properties like excitation energies and excited states. This demonstrates how quantum computing can provide more accurate modeling of molecular behavior compared to classical approximation. (d): Quantum-classical hybrid optimization framework for clinical trial management, where quantum circuits generate trial site configurations while working in conjunction with classical optimizers to calculate loss functions. The framework optimizes across multiple parameters, including enrollment rates and costs, visualized in a matrix of trial sites with varying performance metrics.}
    \label{fig:workflow}
\end{figure*}

Before exploring quantum computing's role in pharmaceutical innovation, readers should familiarize themselves with fundamental concepts from both domains. Box~\ref{box:quantumgconcepts} introduces key quantum methods that enable computational advantages across the drug discovery pipeline, from molecular simulation to secure data handling. Box~\ref{box:drugconcepts} provides essential terminology in drug discovery and development, establishing context for understanding quantum applications in pharmaceutical research. We also summarize the conventional pipeline in Figure~\ref{fig:drugpipline} and offer a visual overview of the entire drug development process in Figure~\ref{fig:workflow}, highlighting strategic points where quantum computing could potentially transform traditional approaches. These resources provide the necessary foundation for appreciating how quantum technologies might address the computational challenges that currently limit classical methods in drug discovery and development.

\begin{figure*}[!t]
\begin{tcolorbox}[
    colback=olive!5!white,
    colframe=white
]
\refstepcounter{boxcounter} 
\label{box:quantumgconcepts} 
\large \textbf{BOX \theboxcounter} 

\vspace{.5em}
\hrule
\vspace{.5em}

\LARGE \texttt{Key quantum methods in drug discovery}
\normalsize

\vspace{.5em}
\hrule
\vspace{1em}

Quantum computing offers novel approaches for accelerating drug discovery through enhanced molecular simulation, optimization, and secure data handling. Here we highlight several fundamental quantum methods:

\textbf{Quantum Phase Estimation (QPE):} QPE determines the eigenvalues of a quantum operator by preparing an initial state, applying controlled unitary operations, and extracting the phase information via an inverse quantum Fourier transform. It provides an exponential speedup over classical methods for molecular energy calculations, crucial for predicting stability and binding affinity. By leveraging quantum Fourier transforms and phase kickback techniques, QPE enables precise modeling of complex biomolecular interactions, which classical diagonalization struggles to achieve efficiently.

\textbf{Variational Quantum Eigensolver (VQE):} VQE is a hybrid quantum-classical algorithm where a quantum circuit prepares trial states, and a classical optimizer updates parameters to minimize energy. It improves upon classical approaches in solving electronic structure problems by leveraging parametrized quantum circuits and hybrid optimization, avoiding the exponential cost of exact diagonalization. With ansätze like Unitary Coupled Cluster and error mitigation techniques, VQE enhances reaction mechanism predictions, particularly for large, strongly correlated systems.

\textbf{Quantum Approximate Optimization Algorithm (QAOA):} QAOA encodes combinatorial optimization problems into an Ising Hamiltonian and optimizes a quantum circuit with alternating cost and mixing operators. A key application in drug discovery is optimizing clinical trial portfolios, which involve high-dimensional data with multiple interdependent parameters. By leveraging quantum interference and entanglement, QAOA efficiently explores complex search spaces, offering faster convergence and improved solutions for trial site selection, patient cohort stratification, and resource allocation.

\textbf{Quantum Generative Adversarial Networks (QGANs):} QGANs consist of a quantum generator and a quantum or classical discriminator, iteratively optimized to generate complex data distributions. They offer a transformative approach to molecular design by generating novel drug-like molecules with optimal properties. QGANs leverage quantum-enhanced probabilistic modeling for more expressive latent representations, enabling efficient sampling from high-dimensional molecular distributions and accelerating lead compound identification beyond classical generative models.

\textbf{Quantum Kernel Methods:} Quantum kernel methods leverage quantum circuits to compute high-dimensional kernel functions used in machine learning models. They enhance classical models by mapping molecular features to quantum Hilbert spaces using parameterized circuits. Unlike classical kernels, quantum kernels exploit entanglement and superposition to encode non-trivial feature relationships efficiently, offering exponential speedups in structure-activity relationship modeling and improving drug-target interaction predictions.

\textbf{Quantum Imaginary Time Evolution (QITE):} QITE simulates imaginary-time evolution by iteratively adjusting quantum circuit parameters to approximate the projection onto the ground state. It accelerates molecular dynamics simulations, particularly for proton transfer reactions and hydrogen bonding networks. QITE provides polynomial speedup in computing transition states and reaction pathways, enhancing accuracy in enzymatic and protein-ligand interactions where classical simulations struggle.

\textbf{Quantum Teleportation:} Quantum teleportation enables the transfer of quantum states between distant qubits via entanglement and classical communication. It ensures secure quantum data transmission, facilitating distributed quantum computing for collaborative pharmaceutical research. Unlike classical methods requiring direct data exchange, quantum teleportation provides theoretically unbreakable security and real-time synchronization of remote quantum processors, ensuring coherence in multi-node quantum simulations for complex drug modeling.

\textbf{Quantum Federated Learning (QFL):} QFL combines quantum computing with federated learning principles to enable privacy-preserving distributed machine learning. It addresses the challenge of integrating patient data securely in drug trials, leveraging quantum homomorphic encryption and differential privacy. Unlike classical federated learning, QFL uses quantum superposition and entanglement to encode data transformations securely, reducing communication overhead while enhancing model convergence and privacy protection in multi-institutional collaborations.

\end{tcolorbox}
\end{figure*}

\section*{A Short Primer on Quantum Computing}
\label{sec:intro} 

The promise of quantum computing in drug discovery stems from a simple fact: molecules are quantum objects. The behavior of electrons and nuclei including bonding, moving and interacting with their environment is governed by quantum mechanics. Classical methods approximate these effects and have achieved remarkable success, but certain phenomena, such as strong electron correlation or subtle protonation dynamics, remain extremely challenging. These quantum details can determine whether a drug binds tightly, selectively, or safely. A computer that works with quantum information has the potential to capture such effects more faithfully, offering new insight where existing tools strain.

What makes quantum computing different is not that it tries “all answers in parallel,” but that it manipulates information in ways that mimic interference and entanglement in nature. The basic unit, the qubit, can exist in a blend of states that can be visualized as a point on a sphere rather than a simple zero or one. When many qubits are combined, they form patterns of correlation that cannot be reproduced by any classical probability model. Quantum algorithms are designed to choreograph these correlations so that the outcome of interest—whether an energy, a probability, or a binding preference—emerges with higher likelihood. In this way, quantum computers extend classical capabilities, but in problem-specific rather than universal ways.

The interest in quantum computing for the simulation of molecules and materials arises because their chemistry and physics are fundamentally quantum mechanical. The state of a many-particle molecular system encodes quantum information. As the number of atoms increases, the description may require an exponentially large number of classical bits. In the worst case, simulating such systems on a classical computer is exponentially hard. This motivates Feynman’s observation that “Nature isn’t classical, dammit, and if you want to make a simulation of nature, you’d better make it quantum mechanical.” A practical quantum advantage in molecular simulation is subtle. Decades of work in electronic structure and quantum chemistry show that classical algorithms can reach reasonable and sometimes very high accuracy with controlled approximations. Any potential advantage is problem specific and must be tied to the question under study and to the accuracy required \cite{bauer2020hierarchy}.

A quantum computer extends classical capabilities by processing quantum information. The basic unit is the qubit, a controllable two-level quantum system with basis states $\ket{0}$ and $\ket{1}$. A general single-qubit state is a superposition:
\begin{align}
    &\ket{\psi} = c_0\ket{0}+c_1\ket{1} \\
    &|c_0|^2+|c_1|^2 = 1.
\end{align}

\begin{figure}[b!]
    \centering
    \includegraphics[width=.6\linewidth]{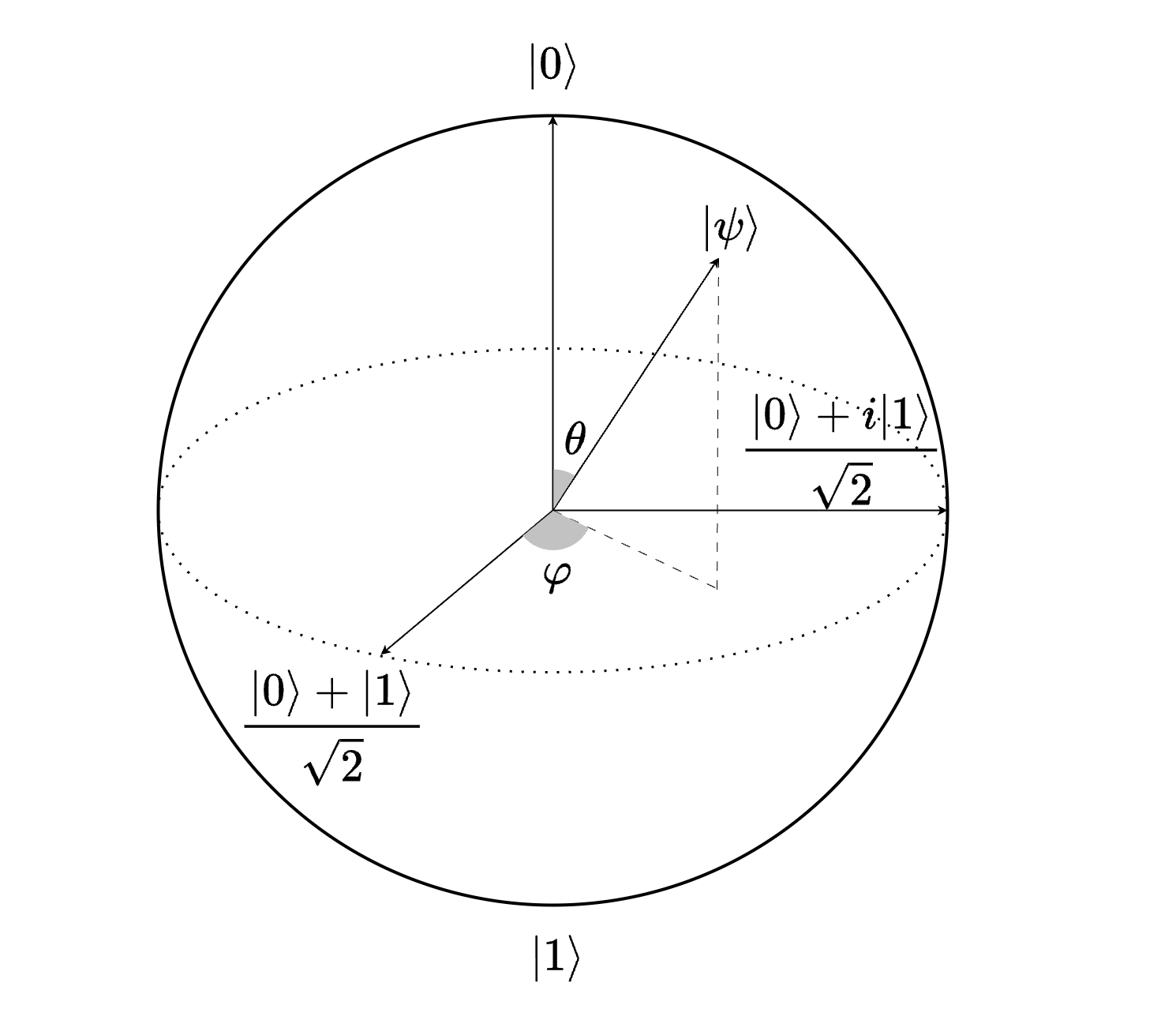}
    \caption{A common way to visualize the state of a single-qubit is to parametrize it as \( |\Psi\rangle = cos(\theta/2)|0\rangle + e^{i\phi}sin(\theta/2)|1\rangle \) where the angles \(\theta\) , \(\phi\) map the state onto a point on the surface of a sphere, known as the Bloch sphere. The north and south poles, \(|0\rangle\) and \(|1\rangle\), represent the “classical states” (or computational basis states) and denote the bits 0, 1 used in a classical computer.}
    \label{fig:bloch}
\end{figure}

This state can be visualized as a point on the surface of the Bloch sphere (Fig.~\ref{fig:bloch}), where the north and south poles correspond to the classical basis states $\ket{0}$ and $\ket{1}$.
For $n$ qubits there are $2^n$ orthonormal basis states $\ket{z_1\ldots z_n}$ with $z_i = 0,1$. A general $n$-qubit state is
\begin{equation}
    \ket{\Psi}=\sum_{z_1,\ldots,z_n = 0, 1} c_{z_1\ldots z_n}\,\ket{z_1\ldots z_n},
\end{equation}
which requires an exponential number of complex amplitudes to specify. Measurement in the computational basis yields a single outcome bitstring $\ket{z_1\ldots z_n}$ with probability $|c_{z_1\ldots z_n}|^2$. Measurements in other bases reveal interference among amplitudes and enable entanglement, which creates correlations that no classical probability distribution can reproduce. These effects underlie the algorithmic power of quantum computing. Quantum algorithms do not gain exponential speedups by trying all answers in parallel. Reading out a quantum state provides only one outcome. Algorithms must therefore choreograph constructive and destructive interference so that the desired outcome appears with high probability and can be extracted with a tractable number of measurements.

\subsection*{Main Quantum Hardware Platforms}
\label{sec:hardware-overview}

Feynman proposed that a controllable quantum system could imitate another quantum system of interest. This inspired analog quantum computation, which has revealed rich physics in cold atom experiments but is limited to specific interactions and lacks systematic error correction. For general algorithms, we focus on digital quantum computation, where qubits are manipulated by quantum gates to form circuits, and on adiabatic/annealing approaches that use quantum dynamics to solve optimization problems~\cite{altman2021quantum, bauer2020quantum}.

Three hardware families currently lead efforts in chemistry and drug discovery: superconducting circuits, trapped ions, and neutral atoms. Table~\ref{tab:hardware} summarizes their strengths, challenges, and relevance.

Superconducting circuits use Josephson devices at millikelvin temperatures, controlled by microwave pulses and read out via resonators. Their advantages include very fast gates and mature control electronics, while challenges include limited connectivity and frequency crowding. For chemistry, their rapid cycle times make them well suited for hybrid algorithms such as variational approaches~\cite{kjaergaard2020superconducting, google2020hartree}.

Trapped ions encode qubits in internal atomic states held by electromagnetic fields. They offer long coherence times and all-to-all connectivity within a trap, though gate speeds are slower and scaling requires modular architectures. Their precision makes them attractive for high-accuracy simulations of small molecules~\cite{bruzewicz2019trapped}.

Neutral atoms arrange qubits in optical tweezers or lattices, with entanglement mediated by Rydberg interactions. They allow flexible geometries and large arrays, but face challenges from atom loss and laser noise. Their tunability offers unique opportunities for mapping molecular structures and for analog simulations of model Hamiltonians~\cite{bluvstein2024logical, henriet2020quantum}.

Across platforms, the control stack now supports calibrated entangling gates, conditional feedforward, and real-time classical processing. Error mitigation techniques are standard, and reproducibility requires reporting device family, connectivity, circuit depth, number of shots, and calibration context. Near-term studies will emphasize shallow circuits on tens to hundreds of physical qubits, while improved fidelities and logical qubits will enable deeper algorithms such as phase estimation in the future~\cite{cai2023quantum, kim2023evidence}.

\subsection*{Practical Scope of Quantum Computation}

In drug discovery, this capability targets fundamental tasks in quantum chemistry and molecular physics. A natural application is real-time quantum dynamics governed by the time-dependent Schrödinger equation, where the Hamiltonian operator $\hat{H}$ encapsulates the total energy of the system:
\begin{equation}
    \mathrm{i}\,\hbar\,\frac{\partial}{\partial t}\ket{\Psi(t)}=\hat{H}\ket{\Psi(t)}, \label{eq:schrodinger}
\end{equation}
from which spectra, response, and dynamical observables follow. Closely related variational formulations target low-energy states($E_0$):
\begin{equation}
    E_0 =\min_{\ket{\Psi}} \bra{\Psi}\hat{H}\ket{\Psi}
\end{equation}
while finite-temperature properties are described using the thermal density matrix($\rho$), which is derived from the partition function($Z$):
\begin{align}
    Z &= \text{Tr}[e^{-\beta \hat{H}}] \\
    \rho &= \frac{e^{-\beta \hat{H}}}{Z} \\
    \langle \hat{A}\rangle_\beta &= \text{Tr}[\rho \hat{A}]. \label{eq:thermo_new}
\end{align}
These formulations govern molecular stability and thermodynamics, providing crucial metrics like binding free energy. Complexity theory offers guidance rather than guarantees. Certain real-time evolutions admit efficient quantum procedures, whereas generic ground-state and thermal problems are QMA-hard in the worst case. QMA is the quantum analogue of NP. In QMA, a quantum verifier checks a proposed quantum proof in polynomial time. A problem is QMA-hard if every problem in QMA can be mapped to it in polynomial time. No efficient quantum algorithm is known for QMA-hard problems. These labels describe worst-case difficulty rather than typical practice. Many chemical instances have structure that algorithms can exploit and some admit controlled approximations. Practical utility therefore depends on how well we prepare relevant states, align the encoding with the physical structure, and control hardware noise and sampling error.

For drug discovery, the goal is a more precise treatment of quantum effects that challenge classical approximations, including electron correlation, polarization, and protonation dynamics, and effects near metal centers. These factors can determine binding, selectivity, potency, and safety. Given the limits of worst-case complexity results, the search for advantage must proceed empirically on chemically relevant instances using real algorithms and devices, with systematic comparison to state-of-the-art classical methods \cite{bauer2020hierarchy}.

Terminology used in this review follows these conventions: a \emph{qubit} is a quantum two-level system manipulated by gates; an \emph{ansatz} is the parameterized circuit used to approximate a state; \emph{active space} is the subset of orbitals treated explicitly on the quantum device; \emph{circuit depth} counts sequential gate layers; \emph{shots} are repeated measurements to estimate expectation values; \emph{error mitigation} denotes post-processing or scheduling that reduces noise bias without full error correction. Throughout, we distinguish hardware-executed demonstrations from simulation-only prototypes and purely theoretical scaling, and we position quantum components as focused evaluators alongside classical pipelines~\cite{mcardle2020quantum, tilly2022variational, cai2023quantum}.

\begin{table*}[ht]
\caption{Comparison of leading quantum hardware platforms for chemistry and drug discovery.}
\label{tab:hardware}
\begin{tabularx}{\textwidth}{l >{\hsize=.9\hsize}X >{\hsize=.8\hsize}X >{\hsize=1.3\hsize}X}
\toprule
\textbf{Platform} & \textbf{Strengths} & \textbf{Challenges} & \textbf{Drug Discovery Relevance} \\
\midrule \vspace{.1cm}
\textbf{Superconducting} & 
Fast single and two-qubit gates; mature control electronics; rapid calibration cycles. & 
Local connectivity; crosstalk and materials loss; frequency crowding. & 
Good for iterative hybrid algorithms like VQE where many short runs are needed. Error-correcting layouts are progressing quickly. \\
\midrule \vspace{.1cm}
\textbf{Trapped ions} & 
Long coherence times; high-fidelity single- and two-qubit gates; uniform qubits. & 
Slower gate speeds; sensitivity to motional heating; scaling requires modular traps and photonic links. & 
Suited for small, high-accuracy simulations of molecular states; all-to-all connectivity within a trap aids chemistry algorithms needing flexible qubit mapping. \\
\midrule \vspace{.1cm}
\textbf{Neutral atoms} & 
Large arrays with flexible geometries; tunable interactions via Rydberg states; support both digital and analog primitives. & 
Atom loss; laser noise and Doppler effects; error correction still at early stage. & 
Flexible layouts may naturally mirror molecular geometries; analog simulation useful for exploring model Hamiltonians relevant to binding and dynamics. \\
\midrule
\end{tabularx}
\end{table*}

\section*{Quantum-driven Drug Discovery}
\label{sec:drug_discovery} 

Drug discovery, beginning with an innovative idea for a new medication, is the initial phase aimed at identifying unique, diverse drug molecules with optimal pharmaceutical attributes. Modern drug discovery and development are highly complex and costly endeavors. Bringing a new drug to market typically requires over a decade and substantial financial investment. In modern drug discovery, computational tools play a crucial role in accelerating the development of new therapies~\cite{hughes2011principles}. These tools help researchers identify potential drug targets, design and optimize molecules, and predict their behavior in the human body.


The molecular interactions underlying drug action are fundamentally quantum mechanical in nature, involving complex electron correlations and conformational dynamics that challenge classical computational methods. Quantum computing offers transformative potential in drug discovery through its native ability to represent quantum states and handle exponential complexity. Understanding how quantum approaches address drug design challenges requires establishing key concepts in both molecular representation and quantum computation.

\subsection*{Foundations of Quantum-Enabled Drug Discovery}
Quantum computing transforms drug discovery by representing molecules in their fundamental quantum mechanical form~\cite{cao2019quantum, motta2022emerging}. Understanding this quantum approach requires examining three distinct representations that enable different aspects of molecular modeling and drug design.

The foundational representation encodes molecular wavefunctions as quantum states in Hilbert space, mapping both electronic and nuclear degrees of freedom to qubit configurations. Unlike classical representations limited to atomic positions and bonds, quantum states capture molecular behavior through superposition and entanglement, naturally accounting for electron correlation effects that classical methods approximate poorly~\cite{ollitrault2021molecular, sun2016quantum, schutt2019unifying, bauer2020quantum}. This direct quantum encoding enables precise modeling of electronic structures and molecular interactions crucial for drug-target binding~\cite{li2024hybrid, sathan2024drug}.

For electronic structure calculations, second quantization provides a practical framework where molecules are described through fermionic operators~\cite{jorgensen2012second}. These operators must be transformed to work with quantum computers' qubit architecture, typically using either Jordan-Wigner or Bravyi-Kitaev encodings~\cite{batista2001generalized, seeley2012bravyi}. These transformations maintain quantum mechanical properties while enabling efficient computation, with Bravyi-Kitaev offering particularly optimized qubit utilization for larger molecular systems~\cite{tranter2018comparison}.

Quantum kernel methods (Box~\ref{box:quantumgconcepts}{}) extend these representations into machine learning applications by mapping molecular structures into high-dimensional quantum feature spaces~\cite{huang2012quantum, zhao2024qksan}. This mapping enables the detection of complex structure-function relationships beyond the reach of classical methods, providing a powerful framework for tasks ranging from binding affinity prediction to lead identification and optimization. These quantum feature spaces bridge pure quantum mechanical representations with practical drug design applications~\cite{batra2021quantum}.

The combination of quantum state representations, fundamental quantum operations, and quantum-enhanced machine learning provides a comprehensive toolkit for addressing the key challenges in structure-based drug design: binding site prediction, binding pose generation, \textit{de novo} ligand generation, binding affinity prediction, and protein pocket generation. In practice we adopt a task-decomposed hybridization: a chemically defined active space at the site of interest is treated with quantum electronic-structure estimators embedded in a classical environment to capture local correlation, polarization, and protonation/spin microstates; discrete design choices are formulated as combinatorial optimizations; and learned surrogates optionally incorporate quantum feature maps/kernels to enrich decision boundaries~\cite{cao2019quantum, tilly2022variational, reiher2017elucidating, mcardle2020quantum, havlivcek2019supervised, schuld2019quantum, cai2023quantum}. 
We apply this template in the sections that follow: for \emph{Protein Binding Site Characterization and Design}, quantum embedding supplies local electronic descriptors at metal centers and catalytic motifs; for \emph{Protein–Ligand Interaction Modeling}, quantum subroutines refine site-specific energetics while docking/diffusion models handle large-scale pose enumeration and ranking; and for \emph{De Novo Ligand Generation}, quantum steps act as physics-aware evaluators on small candidate sets within NISQ limits (with resource reporting: qubits, circuit depth, shots, wall-clock/queue time)~\cite{mcardle2020quantum, cai2023quantum, quek2024exponentially}.

\subsection*{Protein Binding Site Characterization and Design}

Identifying and engineering protein binding sites presents dual challenges: predicting existing binding pockets and designing new ones for targeted drug interactions. Classical approaches using geometric analysis and machine learning methods like MaSIF~\cite{gainza2020deciphering} have advanced our understanding of surface cavities~\cite{senior2020improved, jumper2021highly}, and have substantially improved cavity detection and residue-level descriptors. However, binding sites, particularly in metalloproteins and catalytic centers that are governed by subtle electronic effects, remain difficult for standard force fields and fixed-charge models. This difficulty is well documented by the need for specialized docking force fields and geometry constraints around metal centers (e.g., Zn$^{2+}$)~\cite{santos2014autodock4zn, bai2015accurate}.

Metalloproteins are common as roughly one-third of proteins require or bind metals~\cite{capdevila2024bacterial,lin2024mespeus,durr2023metal3d}, and their coordination chemistry, polarization, and variable protonation/spin can confound fixed-charge scoring; we therefore use them as an \textit{illustrative stress-test}, not a claim of dominance. Quantum methods capture local correlation and protonation/spin microstates via explicit electronic Hamiltonians~\cite{mcardle2020quantum}, but on present hardware are limited to small active spaces with error mitigation~\cite{cai2023quantum,o2023purification,quek2024exponentially}, so we adopt them as complementary evaluators at selected sites while classical tools handle geometry mapping and large-scale search.

Quantum computing can addresses these challenges through Hamiltonian formulation with $\mathcal{O}(N^4)$ scaling~\cite{robert2021resource, doga2024perspective}, enabling more explicit characterization of local electronic effects in small active spaces.
Digital quantum algorithms have been used to estimate relative protonation preferences and indicators of local polarizability~\cite{chandarana2023digitized, cao2018potential}, with potential relevance to metalloproteins such as zinc proteases where electronic structure strongly influences binding. In this setting, quantum subroutines can capture subtle electronic changes that classical fixed-charge force fields must approximate, with proof-of-concept studies reported for catalytic motifs in viral proteases and metalloenzymes~\cite{battaglia2024general, blunt2022perspective}.

The same quantum framework extends naturally to pocket engineering, where electronic effects must be carefully controlled to create stable binding interfaces. Hybrid “quantum-inspired immune” search heuristics~\cite{jiao2008quantum, li2020quantum} have been proposed for identifying dynamic or cryptic pockets that emerge via electronic reorganization~\cite{andersson2022quantum, doga2024perspective}. In selected, peer-reviewed studies, researchers performed experiments on quantum hardware to interrogate binding-relevant electronic descriptors such as estimating protonation-state energetics in model fragments related to staphylococcal nuclease and $\alpha$-lactalbumin~\cite{hu2023quantum}, and mapping dissociation behavior in strongly correlated small-molecule testbeds that benchmark pocket-chemistry motifs~\cite{weaving2025contextual}. In contrast, separate fault-tolerant resource analyses for FeMoco outline how, once error-corrected scales are available, coordination chemistry at metal centers could be treated directly on quantum processors~\cite{reiher2017elucidating}. Collectively, these results should be viewed as proof-of-concept demonstrations: current NISQ devices require error mitigation and restrict active-space sizes, so quantum components presently serve as complementary evaluators of local electronic structure while classical methods carry out geometry mapping and large-scale search~\cite{mcardle2020quantum, quek2024exponentially}.

Early hardware experiments have reported progress on complex binding-site descriptors~\cite{battaglia2024general, blunt2022perspective}, and quantum components have been used to inform design proposals for metalloprotein pockets~\cite{andersson2022quantum, doga2024perspective}. At present, these roles appear most promising when electronic effects dominate pocket formation and stability (e.g., catalytic sites and metal-binding centers). Within current NISQ constraints under limited active-space sizes and the need for error mitigation, we view quantum methods as complementary evaluators of local electronic structure, alongside classical tools that handle geometry mapping and large-scale search.

\subsection*{Protein-Ligand Interaction Modeling}
Predicting protein-ligand binding modes and affinities remains a central challenge in drug discovery. Classical methods like molecular docking programs excel at geometric matching but struggle with electronic effects~\cite{trott2010autodock, alhossary2015fast}. Predicting protein–ligand binding modes and affinities remains a central challenge in drug discovery. Classical methods like molecular docking programs excel at geometric matching but can struggle with electronic effects~\cite{trott2010autodock,alhossary2015fast}. Learning-based approaches such as EquiBind and DiffDock have advanced pose prediction~\cite{stark2022equibind,corso2022diffdock}, yet fundamental limitations persist in modeling quantum-mechanical interactions that determine binding energetics~\cite{hardikar2024quanta,ollitrault2021molecular}.

General-purpose docking and diffusion-based pose prediction are routinely assessed using community-recognized benchmarks that evaluate scoring and ranking power like CASF-2016/PDBbind~\cite{su2018comparative}, pose prediction accuracy across different protein conformations with CrossDocked2020~\cite{francoeur2020three}, and the physical realism of predicted poses using PoseBusters~\cite{buttenschoen2024posebusters}. Representative classical baselines include AutoDock Vina 1.2.0 and AutoDock-GPU, CNN re-scoring with Gnina, and diffusion-based docking with DiffDock~\cite{eberhardt2021autodock, santos2021accelerating, mcnutt2021gnina, corso2022diffdock}. State-of-the-art models also explore protein conformational awareness, such as DynamicBind and equivariant diffusion SBDD~\cite{lu2024dynamicbind, schneuing2024structure, abramson2024accurate}. However, these classical methods often rely on 'fixed-charge' force fields that approximate atoms as points with static charges~\cite{blunt2022perspective}. This simplification presents challenges in accurately modeling fundamentally quantum-mechanical interactions. These challenges include capturing the complex instantaneous interactions of electron correlation, the molecular distortion from dynamic polarization upon binding, and the subtle energetics of different protonation or spin microstates~\cite{capdevila2024bacterial, durr2023metal3d}. Therefore, instead of replacing these highly developed classical pipelines, we position quantum steps as complementary evaluators. By operating on the same quantum rules that underlie chemistry, quantum subroutines can directly represent molecular quantum states to account for these subtle electronic effects~\cite{santagati2024drug, bauer2020quantum, mcardle2020quantum}. They can provide targeted, physics-aware refinement for specific cases where these quantum interactions are critical to binding, addressing a known limitation of otherwise powerful classical tools.

Quantum computing addresses such cases with Hamiltonian-based evaluators over small active spaces. Near-term algorithms (e.g., VQE/QITE, BOX~\ref{box:quantumgconcepts}) can probe local correlation, protonation/spin microstates, and polarization, whereas phase estimation and amplitude estimation primitives target fault-tolerant regimes~\cite{mcardle2020quantum,cai2023quantum,o2023purification,quek2024exponentially,brassard2002quantum}. In practice, we use quantum components to refine local electronic descriptors at selected sites, while classical pipelines handle large-scale pose enumeration, scoring, and ranking~\cite{trott2010autodock,alhossary2015fast,stark2022equibind,corso2022diffdock}.

Recent hybrid quantum-classical frameworks demonstrate practical success in combined pose and affinity prediction~\cite{li2021quantum, uzzaman2019molecular}. These approaches balance computational accuracy with current hardware limitations, as demonstrated in successful predictions for $\beta$-secretase inhibitors~\cite{battaglia2024general, blunt2022perspective}. The integration with molecular dynamics has further enhanced prediction of conformational changes during binding~\cite{ollitrault2021molecular, fedorov2021ab}. Within today’s NISQ regime, applicability is target- and pocket-dependent and largely confined to small active spaces with error mitigation~\cite{mcardle2020quantum, cai2023quantum, o2023purification, quek2024exponentially}; we therefore use quantum steps as complementary evaluators of local electronic descriptors, while classical pipelines handle large-scale pose enumeration, scoring, and ranking. Standardized, hardware-run quantum benchmarks at community scale are not yet available, so reported gains are interpreted as pilot-scale and we accompany any accuracy metrics with resource reporting including qubits, circuit depth, shots, and wall-clock/queue time.

Standardized, hardware-run quantum benchmarks at CASF-2016, CrossDocked2020, or PoseBusters scale are not yet available. To enable standardized and direct comparison we outline: (i) datasets—CASF-2016 (docking/scoring power), CrossDocked2020 (cross-docking), PoseBusters (physical plausibility); (ii) metrics—Top-1/Top-N RMSD (<2~\AA), AUROC for screening, correlation/MAE for affinity, and PoseBusters pass rates; (iii) classical baselines—Vina~1.2 (and AutoDock-GPU), Gnina, DiffDock; and (iv) quantum reporting—physical qubits, circuit depth, shots, wall-clock/queue time, and error mitigation~\cite{su2018comparative, francoeur2020three, buttenschoen2024posebusters, eberhardt2021autodock, santos2021accelerating, mcnutt2021gnina, corso2022diffdock}.

\begin{figure*}[!t]
\begin{tcolorbox}[
    colback=olive!5!white,
    colframe=white
]
\refstepcounter{boxcounter} 
\label{box:drugconcepts}
\large \textbf{BOX \theboxcounter} 
\vspace{.5em}
\hrule
\vspace{.5em}

\LARGE \texttt{Key concepts in drug discovery}
\normalsize
\vspace{.5em}
\hrule
\vspace{1em}
Quantum computing offers novel approaches for accelerating drug discovery through enhanced molecular simulation, optimization, and secure data handling. Here we highlight several fundamental quantum methods enabling these advances:

\textbf{ADMET:} ADMET refers to basic small-molecule drug pharmacokinetics properties, including absorption, distribution, metabolism, excretion, and toxicity.

\textbf{Amino acids:} Amino acids are building blocks for proteins, which come in 20 different kinds. They can combine to form proteins.

\textbf{Antibody:} Antibody refers to a kind of protein that could recognize and bind to the antigen and treat the disease. Antibodies are symmetric and have two identical heavy chains and two identical light chains. 

\textbf{Clinical trial:} A clinical trial is a research study designed to evaluate the safety and efficacy of new medical treatments, drugs, or devices in human participants. These trials follow strict protocols and are conducted in phases, from initial small-scale studies to assess safety to larger-scale studies that confirm effectiveness and monitor side effects.

\textbf{Clinical trial phases:} A drug candidate needs to pass three phases of clinical trials before being approved, where phase I focuses on patient safety, phase II and III investigate safety and efficacy on a larger patient population.

\textbf{\textit{De novo} drug design:} \textit{De novo} drug design refers to a wide category of drug design methods that identify novel and diverse drug molecule structures with desirable pharmaceutical properties from scratch. 

\textbf{Drug repurposing (a.k.a. drug repositioning or drug reuse):} Drug repurposing reuses the approved drugs (or the drugs that pass Phase I) to treat new diseases. 

\textbf{Epitope:} An epitope, or antigenic determinant, is part of a pathogen that antibodies recognize, triggering an adaptive immune response. 

\textbf{Genome-wide association studies (GWAS):} Genome-wide association studies (GWAS) is an observational study that identifies the association between genotypes (genome info) and phenotypes (e.g., trait, disease). 

\textbf{Interventional trial:} Interventional trial is a type of clinical trial where participants will receive intervention/treatment, and the effectiveness of the intervention is evaluated. 

\textbf{Lead compound:} Lead compound refers to the promising molecules that are selected from virtual screening or \textit{de novo} drug design, which is usually not ideal in all the pharmaceutical properties and would be further optimized in lead optimization. 

\textbf{Macromolecule drugs:} Therapeutic molecules with large molecular weight ($>$1000 daltons), e.g., antibodies, enzymes, a carbon atom is 12 daltons, also known as biologics. 

\textbf{Protein structure prediction (a.k.a. protein folding)}: Protein structure prediction, also known as protein folding, predicts the protein 3D structure given the protein amino acid sequence. 

\end{tcolorbox}
\end{figure*}

\subsection*{\textit{De Novo} Ligand Generation}
Desired outcomes for de novo generators include chemical validity, novelty, uniqueness, synthesizability, drug-likeness, targetability, and multi-objective property/ADMET profiles; community benchmarks such as GuacaMol and MOSES define standard tasks and metrics~\cite{brown2019guacamol, polykovskiy2020molecular}. The generation of novel drug molecules presents unique computational challenges, with traditional methods limited by the vast chemical space of approximately $10^{60}$ possible compounds. While geometric deep learning methods like 3DSBDD and Pocket2Mol have advanced structure-based generation through graph neural networks and diffusion models~\cite{luo20213d, peng2022pocket2mol}, they can struggle in specific motifs to capture quantum mechanical properties crucial for binding. Classical approaches particularly falter in designing molecules with specific electronic properties required for target interaction~\cite{bohacek1996art, outeiral2021prospects}. State-of-the-art classical baselines achieve strong benchmark performance and high-throughput inference, including policy-optimized generators (REINVENT) and recent 3D diffusion/equivariant SBDD and conformation-aware models~\cite{blaschke2020reinvent, schneuing2024structure, lu2024dynamicbind}. Classical approaches particularly falter in select cases when designing molecules with electronic properties tightly coupled to target interactions~\cite{bohacek1996art, outeiral2021prospects}.

Quantum algorithms can offer complementary tools transformative solutions through the QAOA framework~\cite{farhi2014quantum, smart2021quantum}. This approach enables efficient exploration of conformational space while maintaining quantum mechanical accuracy, particularly excelling at optimizing electronic properties crucial for binding. Recent adaptive variational algorithms enhance these capabilities through systematic quantum circuit growth~\cite{grimsley2019adaptive, motta2022emerging}, enabling more efficient representation of complex molecular systems.

Recent hybrid quantum–classical optimization approaches~\cite{grimsley2019adaptive, li2024efficient} report pilot-scale designs that balance electronic properties with geometric constraints and are most relevant when electronic effects dominate binding like selected metalloproteins and enzyme active sites.~\cite{andersson2022quantum, liu2022prospects}; within today’s NISQ regime we therefore deploy quantum components as physics-aware evaluators for small candidate sets with resource reporting for high-throughput generation, and we note that standardized, hardware-run quantum leaderboards for de novo generation are not yet available~\cite{mcardle2020quantum, cai2023quantum, quek2024exponentially}.

\subsection*{Language Models for Enhanced Quantum Drug Discovery}
The integration of language models with quantum computing introduces a transformative approach to drug discovery by enhancing both molecular representation and quantum simulation capabilities~\cite{daley2022practical, chen2021quantum}. Modern chemical language models have demonstrated remarkable success in learning complex molecular patterns and generating novel lead compound with desired properties~\cite{moret2023leveraging, grisoni2023chemical}, while quantum computing’s specific advantages for modeling electronic interactions include: (i) explicit treatment of electron correlation in small active spaces via variational estimators, (ii) quantum feature maps/kernels that embed descriptors in high-dimensional Hilbert spaces for non-linear decision boundaries, and (iii) amplitude-estimation–based estimators that can, in principle, reduce sampling requirements for certain expectation values under low-noise assumptions~\cite{mcardle2020quantum, havlivcek2019supervised, schuld2019quantum, brassard2002quantum, cai2023quantum}. The convergence of these technologies offers unprecedented opportunities for accelerating drug development through more accurate prediction of molecular behavior and drug-target interactions~\cite{liu2021ai, chakraborty2023artificial}.

Language models have revolutionized \textit{de novo} drug design through their ability to learn and generate valid molecular structures while maintaining chemical feasibility~\cite{moret2023leveraging, grisoni2023chemical}. These models can effectively capture complex structure-activity relationships and predict bioactivity patterns, particularly when enhanced with quantum mechanical principles. Recent advances in quantum language models have demonstrated superior capabilities in representing molecular states through quantum superposition and entanglement~\cite{coecke2020foundations, meichanetzidis2020quantum, chen2021quantum}, enabling more accurate modeling of electronic structures and binding interactions than classical approaches~\cite{ghazi2025quantum}. The quantum-enhanced language models show particular promise in predicting protein-ligand binding modes and optimizing lead compounds~\cite{mukesh2024qvila, liang2023unleashing}.

The synergy between quantum computing and language models extends beyond simple combination. Quantum language models with entanglement embedding have shown remarkable improvements in capturing non-classical correlations within molecular systems~\cite{chen2021quantum, xie2015modeling}, while quantum-enhanced neural networks have demonstrated superior performance in predicting complex chemical properties~\cite{jia2019quantum}. These advances have enabled more efficient exploration of chemical space by combining the pattern recognition capabilities of language models with the quantum mechanical accuracy of quantum computing~\cite{panahi2019word2ket}. For instance, quantum-infused vision-language models have shown particular promise in understanding molecular interactions through multi-modal representations~\cite{mukesh2024qvila}.

Looking toward practical applications, the integration of language models with quantum computing offers several promising directions for drug discovery. This combined approach enables more accurate prediction of drug-target interactions by incorporating both quantum mechanical effects and learned chemical patterns. The hybrid framework can guide quantum simulations toward promising regions of chemical space~\cite{lanyon2010towards, cao2019quantum}, while language models can help interpret and translate quantum computational results into actionable insights for drug design~\cite{mukesh2024qvila}. As both quantum hardware and language model capabilities continue to advance, this integrated approach promises to significantly accelerate the drug discovery process by combining the complementary strengths of both technologies.

\section*{Quantum-enhanced Clinical Trials}

Drug development refers to the use of clinical trials to evaluate the safety and effectiveness of drug-based treatments on human bodies. It is designed to ensure that innovative drugs are safe, effective, and available to patients in the shortest possible time. Clinical trials represent a critical, resource-intensive bottleneck in drug development, requiring large patient cohorts, coordinated logistics, and precise outcome predictions. Quantum computing can contribute to several facets of trial design and execution, from optimizing site selection to guiding patient recruitment strategies, while maintaining data privacy and security.

\subsection*{Quantum-optimized Clinical Trial Portfolio Management and Site Selection}
The optimization of clinical trial management presents a multifaceted challenge amenable to combinatorial optimization, including classical and quantum approaches, particularly through methods initially developed for quantum portfolio optimization in finance~\cite{rebentrost2024quantum, buonaiuto2023best}. In clinical trial planning, selecting the right sites is treated as a complex optimization problem that must balance large amount of complex practical factors among ensuring broad geographic coverage, achieving high patient recruitment rates, engaging investigators with strong relevant expertise, meeting regulatory timelines for site activation, and controlling costs~\cite{doga2024towards}. An adaptive trial design further complicates planning: this design framework allows certain trial parameters to be adjusted over time based on interim results, according to pre-specified rules, so that the trial can become more efficient and ethical as it progresses~\cite{pallmann2018adaptive}. To solve these multifaceted design and site-selection challenges, researchers have traditionally relied on classical optimization methods such as mixed-integer programming and metaheuristic algorithms~\cite{inan2020digitizing}. Mixed-integer programming (MIP) provides a rigorous mathematical approach to selecting optimal sets of trial sites or design options under multiple constraints, essentially treating decisions like whether to open a given site as binary variables in an optimization model~\cite{bragin2022surrogate}.

When exact formulations cannot be solved at scale, planners use heuristic search. Classical metaheuristics include genetic algorithms and simulated annealing. These methods explore many candidates and return near optimal site portfolios under cost, access, and timeline constraints~\cite{ala2021optimization, evans2023operations}. Quantum optimization methods extend this toolkit. Quantum annealing maps the portfolio choice to an Ising model and uses quantum transitions to escape local minima~\cite{ikeda2019application, carugno2022evaluating, perez2024solving, weinberg2023supply}. The Quantum Approximate Optimization Algorithm(QAOA) alternates problem and mixer steps to increase the chance of high quality portfolios~\cite{he2023alignment, cheng2024quantum}. In practice these quantum methods run inside hybrid workflows that combine classical and quantum steps~\cite{weinberg2023supply, quinton2025quantum}. The workflow proposes candidates, rescores them under operational rules, and iterates. These quantum algorithms have shown case-dependent benefits in prototypes in handling complex multi-parameter optimization problems in financial markets~\cite{orus2019quantum, ding2024coordinating}, with potential direct applications to clinical trial management. Nature's validation of quantum annealing for complex optimization problems~\cite{mott2017solving, venturelli2019reverse}, as demonstrated in particle physics applications, provides strong evidence for its potential in healthcare optimization. On hardware, quantum annealing has been exercised on real scheduling/logistics tasks using D-Wave systems within hybrid solvers, indicating feasibility but not yet end-to-end superiority for clinical site selection~\cite{ikeda2019application, venturelli2015quantum}.

Recent implementations of quantum portfolio optimization techniques have shown particular promise when adapted to clinical trial management~\cite{doga2024towards, doga2024can, moradi2022clinical}. By encoding trial parameters into quantum Hamiltonians using techniques refined in financial applications, these approaches enable simultaneous optimization across multiple constraints including geographical distribution, patient demographics, and site capabilities~\cite{elsokkary2017financial, bhasin2024enhancing}. The integration of reverse quantum annealing protocols, which have shown over 100-fold speedup in financial portfolio optimization, offers promising applications for dynamic trial management~\cite{venturelli2019reverse}. These methods are particularly effective when initialized with classical solutions and then refined through quantum optimization.

Quantum-enhanced resource allocation has been demonstrated in pilot healthcare settings particularly in scenarios requiring real-time optimization and decision-making~\cite{bhatia2020quantum, duong2022quantum}. These approaches leverage quantum parallelism to simultaneously evaluate multiple trial configurations, with recent implementations showing particular promise in managing distributed healthcare resources~\cite{veni2022quantum}. The integration of quantum deep reinforcement learning has further enhanced these capabilities, enabling adaptive optimization of resource allocation in response to changing trial conditions~\cite{niraula2021quantum, ullah2024quantum, rani2023quantum}.

The practical implementation of these quantum approaches benefits significantly from hybrid quantum-classical frameworks~\cite{raparthi2021real, ho2022quantum}, which combine quantum optimization for complex decision-making with classical systems for data management and regulatory compliance~\cite{sundaram2025challenges, jeyaraman2024revolutionizing}. These hybrid systems have shown particular promise in addressing healthcare-specific challenges, including regulatory requirements and patient privacy concerns. Recent advances in quantum affective processing for multidimensional decision-making have further enhanced the capability to balance multiple competing objectives in trial management.

\subsection*{Quantum Machine Learning for Cohort Stratification and Patient Recruitment}
Quantum machine learning approaches have been explored as complements to classical ML in patient cohort identification, with recent validations showing unprecedented advantages in analyzing complex quantum states that parallel the multidimensional nature of patient data~\cite{motta2020determining, cao2019quantum, liu2021rigorous}. The quantum advantage in handling high-dimensional data spaces, first demonstrated in particle physics applications~\cite{mott2017solving}, has particular relevance for clinical trial recruitment, where the integration of genomic, clinical, and imaging data creates computational challenges that exceed classical capabilities~\cite{kirsopp2022quantum, durant2024primer}.

Recent implementations of quantum kernel methods have shown remarkable efficiency in mapping data into higher-dimensional Hilbert spaces, where clinically relevant patterns become more distinguishable~\cite{chen2021quantum, narottama2021quantum}. These approaches, building on quantum state manipulation techniques, enable simultaneous analysis of multiple data modalities - from genomic markers to longitudinal clinical records - with a computational efficiency unattainable through classical methods~\cite{mukesh2024qvila, avramouli2024hybrid}. The quantum advantage becomes particularly evident when handling strongly correlated datasets, such as genetic profiles combined with clinical outcomes, where quantum systems can naturally represent the inherent entanglement of biological systems~\cite{ho2022quantum, song2022quantum}. Systematic assessments in digital health find no consistent superiority of QML over strong classical baselines to date~\cite{gupta2025systematic}. Where explored, studies should disclose device type, qubits, circuit depth, shots, error-mitigation, and training/inference wall-clock.

The practical implementation of these quantum approaches leverages recent advances in quantum-classical hybrid architectures~\cite{raparthi2021real, duong2022quantum}. Quantum classifiers can simultaneously process multiple data types while maintaining the coherence necessary for identifying subtle patient subgroups~\cite{ezhov2001pattern, schuld2014quantum}. These hybrid systems have shown particular promise in rare disease trials, where quantum-enhanced pattern recognition enables identification of suitable patients through complex combinations of biomarkers and clinical indicators. The integration of quantum annealing techniques, validated through complex optimization problems~\cite{chen2022optimizing, ikeda2019application, sahner2020clinical}, further enhances the ability to navigate the vast space of patient characteristics efficiently.

Looking toward practical applications, recent Nature-validated quantum implementations have demonstrated significant advantages in processing real clinical data. These systems show particular promise in personalizing trial recruitment strategies through quantum-enhanced analysis of patient-specific factors~\cite{sundaram2025challenges}. As quantum hardware capabilities continue to advance~\cite{acharya2024quantum, castelvecchi2024truly}, these approaches offer a path toward more efficient and accurate patient stratification, potentially reducing both the time and cost associated with clinical trial recruitment.

\subsection*{Quantum Federated Learning for Secure and Private Data Integration}

\begin{figure}[t]
    \centering
    \includegraphics[width=\linewidth]{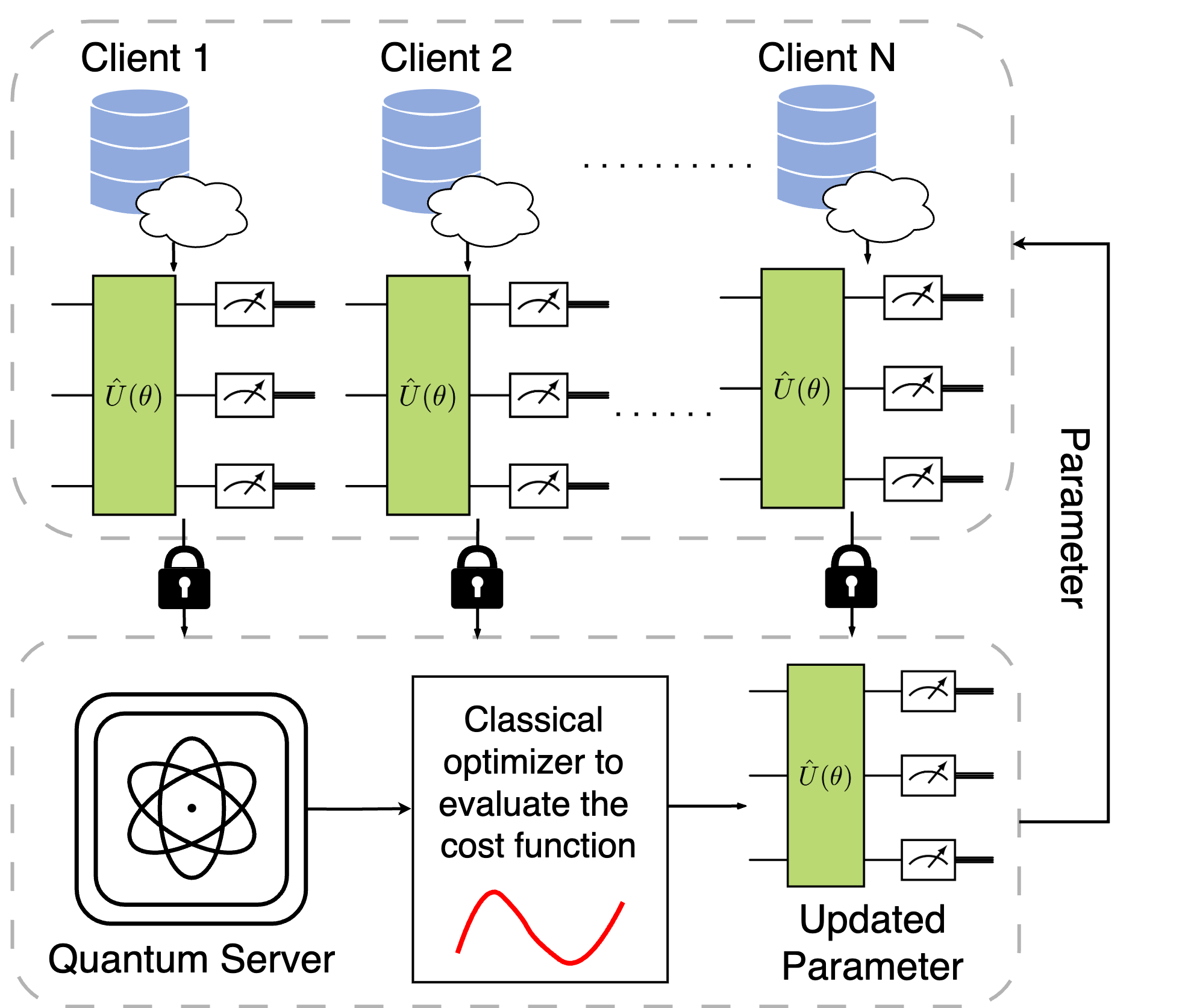}
    \caption{\textbf{Schematic illustration of QFL.}  Multiple clients hold private data that can not be leaked to or shared with the outside, while the QFL scheme enables the training of a public model without uploading their private data.}
    \label{fig:QFL}
\end{figure}

Effective clinical trial design often requires the integration of sensitive patient data from multiple hospitals and research centers. Federated learning addresses privacy concerns by allowing these institutions to collaborate to train global models without sharing raw data~\cite{Goddard2017EU, Li2020Review, Abadi2016Deep}. Each participant optimizes the model parameters locally, transmitting only updates to a central aggregator, thus preserving confidentiality. Quantum federated learning (QFL, BOX~\ref{box:quantumgconcepts}) enhances this paradigm by using QML models to handle complex high-dimensional clinical data more efficiently than classical methods~\cite{Ren2024Quantum, Chehimi2024Foundations, Zheng2023Speeding, Molteni2024Exponential}. In this context, “quantum-federated” variants are best understood as (i) classical FL orchestration with (ii) optional quantum subroutines (e.g., quantum kernels) running locally on privacy-shielded sites, and (iii) end-to-end security aligned with current standards. The intrinsic randomness of quantum mechanics can be integrated with differential privacy(Fig~\ref{fig:QFL}), which naturally provides noise for stronger privacy guarantees~\cite{Caro2022Generalization, Chehimi2022Quantum}. Recent advances in multimodal QFL with fully homomorphic encryption (FHE) have demonstrated particular promise in maintaining model performance while ensuring data privacy, especially when handling diverse medical data types such as combined genomics and imaging data~\cite{dutta2024mqfl}.

Quantum ML models show remarkable efficiency in processing high-dimensional clinical data, with multimodal quantum mixture of experts (MQMoE) frameworks demonstrating significant advantages in handling heterogeneous medical datasets~\cite{yu2023mmoe, dutta2024mqfl}, although privacy leakage must be evaluated with rigorous membership-inference and attribute-inference tests. The integration of quantum randomness with differential privacy provides natural noise injection for enhanced privacy guarantees~\cite{watkins2023quantum, hirche2023quantum}, while quantum generative adversarial networks (QGANs, BOX~\ref{box:quantumgconcepts}) enable the creation of synthetic datasets that preserve essential statistical properties while protecting patient identities~\cite{Gao2022Enhancing}. These approaches show particular promise for underrepresented patient categories, addressing a crucial challenge in clinical trial diversity.

The security of clinical trial data benefits significantly from quantum-enhanced detection systems. Quantum kernel methods implemented in Support Vector Machines have achieved detection accuracies of 75-95\% for label-flipping attacks on medical sensor data~\cite{onim2025detection}, substantially outperforming classical approaches. The integration of these quantum attack detection systems with federated learning frameworks creates a robust security infrastructure that can identify subtle manipulation patterns across multiple trial sites.

Advanced quantum cryptographic techniques, including blind quantum computing and post-quantum cryptography~\cite{li2021federal, Broadbent2009Universal, Bernstein2017Postquantum}, enable secure computations on encrypted data while ensuring long-term protection against future quantum threats~\cite{gupta2024intelligent}. The combination of these approaches with multimodal QFL creates a comprehensive framework for secure, privacy-preserving clinical trial data analysis. This hybrid quantum-classical approach enables real-time monitoring of trial data while maintaining computational efficiency and strict privacy standards. In practice, today’s deployments should rely on established FL privacy tools, with quantum components treated as experimental modules whose contribution is measured against strong classical baselines. We therefore treat QFL as prospective; evaluations should report communication rounds, cryptographic overheads, device specs, and ablations versus classical FL baselines~\cite{nguyen2022federated, antunes2022federated}.

\section*{Outlook}
As we assess quantum computing's transformative potential in drug discovery, we have systematically analyzed the computational resource requirements and expected advantages across key application areas (Table~\ref{tab:comp_resources}). Our analysis covers the spectrum from molecular structure simulation to clinical trial optimization, comparing classical computing requirements with both current NISQ-era capabilities and projected fault-tolerant quantum systems~\cite{cao2019quantum, liu2022prospects, yi2024complexity}. While current quantum devices face hardware limitations, projected capabilities of fault-tolerant quantum computers suggest significant speedups across all major drug discovery applications~\cite{wang2023recent, baiardi2023quantum}. These estimates, validated through initial implementations on current quantum hardware, provide a roadmap for achieving practical quantum advantage in pharmaceutical development~\cite{motta2020determining, motta2022emerging}.

\begin{table*}[!h]
\caption{Computational resource requirements comparison for drug discovery applications. }
\label{tab:comp_resources}
\small  
\begin{tabularx}{\textwidth}{>{\raggedright\arraybackslash}p{1.9cm}>{\raggedright\arraybackslash}p{3cm}>{\raggedright\arraybackslash}X>{\raggedright\arraybackslash}X>{\raggedright\arraybackslash}p{5.2cm}}
\toprule
\textbf{Application Area} & \textbf{Classical Computing} & \textbf{Current Quantum Computing (NISQ)} & \textbf{Future Quantum Computing (Fault-Tolerant)} & \textbf{Expected Advantage} \\
\midrule

Molecular Structure Simulation & 
$10^3$-$10^6$ CPU hours for complex proteins; Requires supercomputing clusters; Memory usage: 100GB-1TB~\cite{paul2010improve, kitchen2004docking} & 
Limited to small molecules ($<$10 atoms); 50-100 noisy qubits; Runtime: hours to days; High error rates limit accuracy~\cite{babbush2018low, mcardle2020quantum} & 
Estimated $10^3$-$10^4$ logical qubits; Minutes to hours runtime; Negligible error rates through error correction~\cite{cao2019quantum, babbush2018low} & 
Exponential to polynomial advantage: Classical $\mathcal{O}(e^N)$ to Quantum $\mathcal{O}(N^4)$ for electron calculations, enabling simulation of large molecules intractable on classical computers; QPE: exponential speedup; VQE: $\mathcal{O}(poly(N))$ ground state calculation~\cite{jiang2021survey, yuan2020quantum, tilly2022variational, motta2022emerging}; QITE: $\mathcal{O}(poly(N))$ proton transfer simulation~\cite{motta2020determining}. \\
\midrule

Property Prediction (ADMET) & 
Training: hours–days (1–8 GPUs); Inference: ms–s/molecule, $\sim 10^5$–$10^6$/day/GPU; web tools available~\cite{daina2017swissadme, xiong2021admetlab} & 
Currently impractical for real molecules; Requires 100-500 noisy qubits; Limited by decoherence~\cite{kandala2017hardware} & 
500-1000 logical qubits; Hours runtime; Quantum machine learning advantages~\cite{mcardle2020quantum, moll2018quantum} & 
Quadratic advantage: classical $\mathcal{O}(N)$ to $\mathcal{O}(\sqrt{N})$ in certain classification tasks; Quantum Kernels: $\mathcal{O}(log N)$ feature mapping; QGANs: classical $\mathcal{O}(exp(N))$ sampling to $\mathcal{O}(poly(N))$~\cite{ciliberto2018quantum, cerezo2022challenges, rebentrost2014quantum}.\\
\midrule

Drug-Target Interaction & 
CPU/GPU: seconds–minutes/complex; ML docking often minutes; AF3 minutes/complex; HPC for large cross-docking~\cite{eberhardt2021autodock, santos2021accelerating, corso2022diffdock, abramson2024accurate, bender2021practical} & 
Not yet feasible for protein-scale interactions; Requires thousands of qubits~\cite{cao2019quantum} & 
$10^4$-$10^5$ logical qubits; Days runtime; Full quantum simulation of binding dynamics~\cite{mcardle2020quantum, pyrkov2023quantum} & 
Exponential to polynomial advantage: Classical $\mathcal{O}(e^N)$ to Quantum $\mathcal{O}(N^4)$ for molecular dynamics simulation; Particularly significant for large protein-ligand systems~\cite{austin2012quantum}; QPE: exponential speedup binding energies; QITE: for H-bonds classical $\mathcal{O}(exp(N))$ to $O(N²)$; QAE: classical $\mathcal{O}(N)$ to $\mathcal{O}(\sqrt{N})$ in sampling~\cite{motta2020determining, plekhanov2022variational}. \\
\midrule

Clinical Trial Optimization & 
Hours to days on distributed systems; Limited by classical optimization algorithms~\cite{scannell2012diagnosing, fogel2018factors, turon2024infectious} & 
100-200 qubits for small-scale optimization; Minutes to hours runtime~\cite{reiher2017elucidating, moll2018quantum} & 
500-1000 logical qubits; Minutes runtime; Quantum optimization advantage~\cite{moll2018quantum} & 
QAOA: Polynomial advantage for combinatorial optimization problems also provide speedup scaling with problem size for trial design optimization; QFL: classical $\mathcal{O}(N)$ to $\mathcal{O}(log N)$ in communication; Quantum Teleportation: $\mathcal{O}(1)$ secure transfer~\cite{guerreschi2019qaoa, zhou2020quantum, venturelli2019reverse}.\\
\midrule

Lead Discovery(vHTS/Hit Identification) & 
$10^6$–$10^8$ screened in hours–days (cloud/GPU); ultra-large runs demonstrated~\cite{lyu2019ultra, gorgulla2020open, gentile2022artificial} & 
Currently limited to proof-of-concept; 100-500 noisy qubits~\cite{cao2019quantum, moll2018quantum} & 
1000-5000 logical qubits; Days runtime; Quantum search advantage~\cite{cao2019quantum, pyrkov2023quantum} & 
Quadratic advantage: $\mathcal{O}(\sqrt{N})$ from $\mathcal{O}(N)$ for chemical space search using quantum search algorithms; Advantage increases with size of chemical library~\cite{grassl2016applying, lloyd2014quantum}: QGANs \& Adaptive Algorithms: classical $\mathcal{O}(exp(N))$ to $\mathcal{O}(poly(N))$~\cite{jain2022hybrid}; Quantum language models: estimation $\mathcal{O}(\sqrt{N})$ in sampling.\\
\bottomrule[0.7pt]
 \end{tabularx}

\begin{tablenotes}
\small
\item Notes: 
\item (1) Classical computing metrics are based on state-of-the-art systems and methods
\item (2) NISQ estimates reflect current hardware limitations and error rates
\item (3) Fault-tolerant projections assume mature error correction and quantum memory
\item (4) Expected advantages are theoretical and depend on algorithm development
\item (5) Resource requirements may vary significantly based on problem specificity and desired accuracy
\item (6) In the NISQ regime, cost-effective quantum use arises primarily as selective, physics-aware evaluators that reduce experimental iterations, consistent with recent hardware-executed chemistry/design pipelines and evidence for pre–fault-tolerant utility~\cite{kim2023evidence, maskara2025programmable, ghazi2025quantum, zhang2025prediction}

\end{tablenotes}
\end{table*}

In the near term, quantum workloads that fit drug‐discovery practice are best scoped as complements to classical compute: small active‐space electronic descriptors (e.g., local correlation/protonation indicators), variational or imaginary‐time estimators, kernel‐based feature maps in niche classification regimes, and modest combinatorial encodings. Execution is predominantly via cloud services with metered billing, while a limited number of sites operate on-prem systems for health/academic research (e.g., IBM Quantum System One installations at Cleveland Clinic, Rensselaer Polytechnic Institute, and Yonsei University)\,\cite{murphy2025ibm, steffen2011quantum}. Other universities maintain in-house research platforms without procuring a vendor system like trapped-ion at Duke and superconducting programs at Yale. These access models make pilot-scale studies feasible for medium-sized institutions and SMEs via pay-as-you-go or committed-minute plans, with careful scoping to physics-aware evaluation that can reduce downstream synthesis/assay burden rather than replace high-throughput docking/ADMET. In practice, adopters already range from large pharma to startups: Boehringer Ingelheim has an active collaboration with Google Quantum AI~\cite{arute2019quantum}; Roche’s pRED unit has run quantum-ML explorations with QC Ware~\cite{mathur2021medical}; Qubit Pharmaceuticals (SME) is executing drug-discovery tasks on PASQAL’s neutral-atom QPUs~\cite{d2024leveraging}; and the start up company Menten AI used D-Wave’s hybrid solver to generate and synthesize peptide designs. Ecosystem pilots also include AstraZeneca working with IonQ via AWS/NVIDIA~\cite{zhao2025quantum}. This positioning aligns with recent evidence of pre–fault-tolerant “utility” under error mitigation\,\cite{kim2023evidence} and hardware-executed chemistry/design pipelines\,\cite{maskara2025programmable, ghazi2025quantum, zhang2025prediction}.

Hardware‐relevant, drug‐adjacent signals are beginning to appear across the literature. A \emph{Nature Biotechnology} study reports a quantum–computing–enhanced generative pipeline that proposed KRAS inhibitors, with 15 molecules synthesized and two showing promising activity~\cite{ghazi2025quantum}. Reconfigurable quantum processors have executed programmable simulations of correlated spin/electronic models with chemically relevant spectral extraction~\cite{maskara2025programmable}. Drug-facing building blocks include a hardware-executable framework for protein 3D structure prediction~\cite{zhang2025prediction} and a quantum algorithm for protein side-chain optimisation with explicit comparisons to classical baselines~\cite{agathangelou2025quantum}. On the learning side, quantum kernel methods have been applied to drug–target interaction prediction in recent~\cite{pallavi2025qkdti}. In parallel, QAOA/DC-QAOA formulations for docking are being developed and stress-tested in simulation and small-scale runs~\cite{ding2024molecular, papalitsas2025quantum}. More broadly, pre–fault-tolerant “utility” under error mitigation has been documented on large devices~\cite{kim2023evidence}.

The field of quantum computing for drug discovery stands at a transformative moment, with recent advances in both quantum simulation capabilities and artificial intelligence suggesting multiple paths toward practical quantum advantage in pharmaceutical development~\cite{daley2022practical}. Current quantum devices have demonstrated the ability to simulate systems of approximately 100 qubits; however, mapping that scale onto realistic biomolecular systems remains out of reach for NISQ hardware and will require error-corrected, fault-tolerant systems. This achievement points toward the potential creation of a quantum-enabled database of amino acid configurations~\cite{baiardi2023quantum}, which could dramatically accelerate protein structure prediction and drug development processes. Such a database would not only enhance our ability to explore protein conformations but could also serve as a benchmark to demonstrate quantum advantage over classical methods in molecular modeling~\cite{ollitrault2021molecular}.

The development of a quantum database for protein folding represents a particularly promising direction for near-term applications. While classical databases store static structural information, a quantum database could maintain the full quantum mechanical description of amino acid configurations, including crucial electronic states and quantum correlations that influence protein folding pathways~\cite{roy2013quantum, hamouda2016quantum}. This quantum representation would enable direct simulation of folding dynamics, potentially revealing intermediate states and alternative conformations that are computationally inaccessible to classical methods~\cite{pal2024quantum}. The database could also incorporate quantum error correction protocols to maintain the fidelity of stored quantum states, ensuring reliable access to complex molecular configurations for drug design applications. By systematically cataloging quantum states of amino acid pairs and their interactions, researchers could build up a library of fundamental building blocks for protein structure prediction, potentially enabling the simulation of larger protein segments through the composition of these basic units.

Building upon this quantum database foundation, the emerging synergy between large language models and quantum computing hints at transformative possibilities for molecular simulation. Recent work has revealed unexpected connections between the mathematical structures underlying both quantum systems and language models~\cite{chen2021quantum}, suggesting deeper theoretical links that could be exploited for mutual benefit. Language models could enhance quantum simulations by learning efficient representations of molecular states and guiding the exploration of chemical space~\cite{ross2022large}, while quantum computers could provide unique capabilities for training and optimizing these models~\cite{outeiral2021prospects, baiardi2023quantum, andersson2022quantum}. This cooperative framework could lead to breakthrough approaches in molecular simulation, potentially accelerating drug discovery by combining the complementary strengths of both paradigms.

To fully realize these capabilities at scale, distributed quantum computing architectures present a promising direction for expanding drug discovery applications. This approach naturally aligns with the distributed nature of drug development, where different aspects of molecular design and testing often occur across multiple sites. Quantum teleportation (BOX~\ref{box:quantumgconcepts}) protocols could enable secure information transfer between distributed quantum processors~\cite{ding2024coordinating}, creating a network of smaller quantum systems that collectively function as a larger quantum computer. This distributed architecture offers two significant advantages: it provides a path to scaling up quantum resources beyond single-device limitations~\cite{liu2021distributed, main2025distributed}, and it enables secure data sharing through quantum cryptographic protocols~\cite{polacchi2023multi}.

The integration of federated learning approaches with distributed quantum computing suggests a powerful framework for collaborative drug discovery. Local quantum processors could perform molecular simulations and property predictions while maintaining data privacy through quantum-secured federated learning protocols~\cite{Ren2024Quantum, Chehimi2024Foundations, Zheng2023Speeding}. A central quantum server could aggregate insights from multiple sites without compromising sensitive information, protected by quantum cryptographic principles. This architecture could enable pharmaceutical companies and research institutions to collaborate more effectively while maintaining strict control over their intellectual property and patient data.

As quantum hardware continues to advance, the integration of these complementary technologies - quantum databases, language model enhancement, and distributed quantum computing - could create a powerful ecosystem for drug discovery. The combination of increased simulation scale, intelligent guidance through language models, and secure distributed computation suggests a path toward quantum advantage in practical drug development applications. While significant challenges remain in scaling quantum devices and developing robust distributed architectures, the convergence of these technologies offers promising directions for transforming the drug discovery process.

\section*{Acknowledgement}
We thank Weikang Li for his helpful discussions. JL is supported in part by the University of Pittsburgh, School of Computing and Information, Department of Computer Science, Pitt Cyber, and the PQI Community Collaboration Awards. This research used resources of the Oak Ridge Leadership Computing Facility, which is a DOE Office of Science User Facility supported under Contract DE-AC05-00OR22725.
Z.L.’s work is supported in part by the IBM–RPI Future of Computing Research Collaboration and in part by the National Science Foundation under Award No. 2519029. FTC's work is funded in part by EPiQC, an NSF Expedition in Computing, under award CCF-1730449; in part by STAQ under award NSF Phy-1818914/232580; in part by the US Department of Energy Office of Advanced Scientific Computing Research, Accelerated Research for Quantum Computing Program; and in part by the NSF Quantum Leap Challenge Institute for Hybrid Quantum Architectures and Networks (NSF Award 2016136), in part based upon work supported by the U.S. Department of Energy, Office of Science, National Quantum Information Science Research Centers, in part by Wellcome-Leap’s Q4Bio program, and in part by the Army Research Office under Grant Number W911NF-23-1-0077. The views and conclusions contained in this document are those of the authors and should not be interpreted as representing the official policies, either expressed or implied, of the U.S. Government. The U.S. Government is authorized to reproduce and distribute reprints for Government purposes notwithstanding any copyright notation herein. FTC is the Chief Scientist for Quantum Software at Infleqtion and an advisor to Quantum Circuits, Inc.

\section*{Author Contributions}
Y.Z, JT. C, JL. C, X. C, and Y.Z, wrote the main manuscript, Y.Z and JT.C are prepared figures, G.K, M.Z, and F.C, are observe the project, Z.L, J.L, and T.F are mentored the project. All authors reviewed the manuscript.
\section*{Competing Interests}
The authors declare that they have no competing interests.

\thispagestyle{empty}

\printbibliography

\end{document}